# RF plasma cleaning of optical surfaces: A study of cleaning rates on different carbon allotropes as a function of RF powers and distances

M. González Cuxart,[1] J. Reyes-Herrera,[1] I. Šics,[1] A. R. Goñi,[2,3] H. Moreno Fernandez,[1]

V. Carlino,[4] and E. Pellegrin[1]

[1] ALBA Synchrotron Light Facility, Carretera BP1413, km 3.3,
08290 Cerdanyola del Vallès (Barcelona), Spain

[2] Institut de Ciència de Materials de Barcelona (ICMAB-CSIC), Campus UAB,
08193 Bellaterra (Barcelona), Spain

[3] ICREA, Passeig Lluís Companys 23, 08010 Barcelona, Spain

[4] ibss Group Inc., 111 Anza Blvd., Suite 110, Burlingham CA, 94010, USA

**ABSTRACT**

An extended study on an advanced method for the cleaning of carbon contaminations on large optical surfaces using a remote inductively coupled low–pressure RF plasma source (GV10x downstream asher) is reported in this work. Technical as well as scientific features of this scaled up cleaning process are analysed, such as the cleaning efficiency for different carbon allotropes (amorphous and diamond-like carbon) as a function of feedstock gas composition, RF power (ranging from 30 to 300W), and source-object distances (415 to 840 mm). The underlying physical phenomena for these functional dependences are discussed.

**Keywords:** Remote inductively coupled plasma, Plasma cleaning, Graphitic carbon, Diamond-like carbon, Optical emission spectroscopy, X-ray photoemission spectroscopy, Raman Spectroscopy

1. **INTRODUCTION**

Previous publications in the field of low-pressure radiofrequency (RF) plasma cleaning of optical surfaces have shown that it is possible to efficiently and safely remove carbon contaminations at the molecular scale [1-7] by using $O_2$/Ar or $H_2$/Ar mixtures as plasma feedstock gases. In our earlier study [1], the main motivation was to report on a new, safe, and efficient method with well-defined sets of process parameters capable to perform an in-situ cleaning of carbon-contaminated optical elements that are typically enclosed within UHV chambers, such as synchrotron optics, extreme ultraviolet (EUV) optics, high-power laser optics, SEM/TEM setups, etc. In order to achieve this purpose, we did perform a comparative work using several different direct capacitive coupled (d-CCP) as well as commercial remote inductively coupled (r-ICP) model GV10x plasma sources operating at different RF powers.

In this previous publication [1], we have been working at a maximum RF power of 100W, at a close constant distance of about 308 mm between the plasma source and the test sample to be cleaned, and in a UHV test chamber of limited/compact dimensions. However, the UHV chambers where such in-situ cleaning methods will be employed in the field are generally larger than the test chamber that has been used so far. Thus, moving to a larger test chamber was required as an attempt to approach realistic cleaning conditions. Moreover, a more powerful version of the GV10x plasma source has been used (up to 300 W RF power), which allowed testing the cleaning process at larger source-sample distances and thus to perform a study of cleaning rates as a function of RF plasma power and distances. These data are intended to contribute to the optimization of the cleaning process, thus making it more efficient especially for larger optical/experimental setups.

On the other hand, it is essential to take into account that the real chemical, crystallographic, and morphological structure of carbon contaminations can be rather complex: Depending on the specific case, these carbon traces are not only consisting of one single carbon allotrope, but may have an amorphous-like nature, composed of different contributions from $sp^2$- and/or $sp^3$-hybridized carbon species or, in other words, have either a more graphitic or diamond-like nature, etc. Therefore, in this study we also determined the cleaning rates of amorphous carbon and diamond-like carbon, as a first step towards eventually obtaining a complete library of cleaning rates as a function of the specific carbon chemistry/configuration.

This study is structured as follows: After a concise description of the experimental setup, we present some results from the sample characterization using x-photoemission spectroscopy (XPS) and Raman spectroscopy. In the next chapter, cleaning rates from the experiments carried out with an $O_2$/Ar plasma using different RF powers and at two different plasma source-sample distances are given. Following the same scheme, further results from experiments done under $H_2$/Ar plasma are given. Samples to be cleaned for both studies were amorphous carbon samples (*Amorphous C*). Finally, cleaning rates of diamond-like carbon samples (*DLC*) (as compared to amorphous C samples) using $O_2$/Ar and $H_2$/Ar plasma feedstock gases with different RF powers at a fixed source-sample distance will be presented.

Last but not least, we also compare the performance of a 100 W model GV10x P4 plasma source (29 $cm^3$ plasma tube volume) with respect to a 300 W model GV10x P3 plasma source (44 $cm^3$ plasma tube volume). With the exception of **Fig. 5(b)**, all data shown in this study have been generated using the larger GV10x P3 plasma source.

## 2. EXPERIMENTAL

### 2.1 Experimental setup

The cleaning experiments were carried out in a horizontal 55 liter volume UHV chamber of cylindrical horizontal shape, with the GV10x plasma source installed at one chamber end and the 250 l/s turbo molecular pumping unit at the opposite end, so that a constant flux of chemically active species from the upstream plasma along the longitudinal axis of the chamber was established (see **Fig. 1**). Two different GV10x downstream asher plasma sources have been used: The P4 source with a smaller upstream plasma volume and a maximum RF power of 100 W as well as the P3 source with a larger plasma volume by a factor of 1.5 and a maximum RF power of up to 300 W. Carbon coated test samples were placed in two quartz crystal microbalances (QCMs) downstream the plasma sources at two different source-sample distances (*QCM 1*: 415 mm, and *QCM 2*: 840 mm). Carbon cleaning rates were calculated from the time evolution of the carbon thickness removed by the plasma as measured by these two QCM balances. Please see reference [1] for further experimental details as well as on the diagnostic tools used in the present study.

### 2.2 Sample preparation and characterization

Two chemically different carbon depositions on gold-coated QCM crystals were prepared as the test samples to be cleaned. In one case, an amorphous carbon thin film (i.e., a mixture of $sp^2$ and $sp^3$ carbon) was deposited by e-beam evaporation from graphitic carbon targets (Goodfellow carbon target model C 009600), carried out in a conventional UHV chamber. We label and refer to them as *Amorphous C.* A diamond-like carbon sample thin film (i.e., $sp^3$ carbon) was deposited by an arc-discharge method, labelled as *DLC* throughout this study. Please note that the surfaces of the different carbon-coated QCM crystals (i.e., the amorphous C and DLC test samples) were not directly facing the plasma jet from the source using normal incidence but rather via grazing incidence (see **Fig. 1**, Top View), so that the plasma kinetic effects – if any - were minimized and thus only chemical effects were effective during the cleaning process.

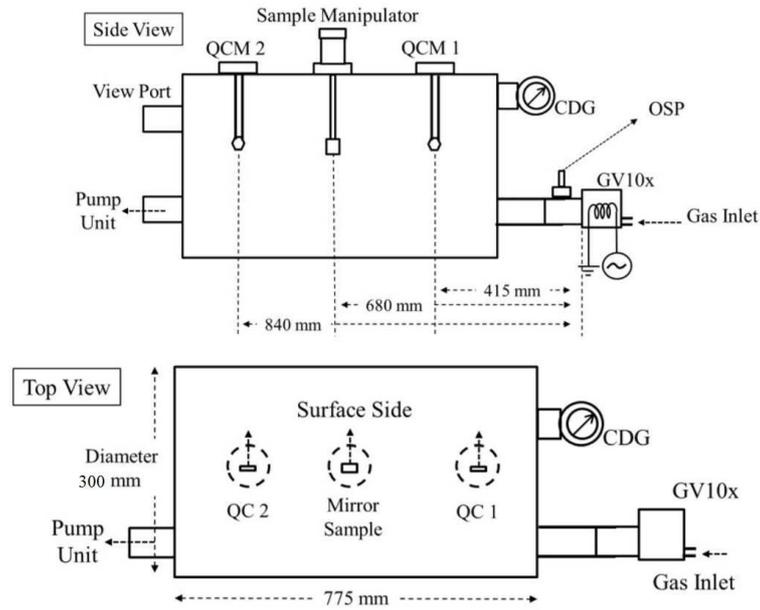

**Fig. 1** Conceptual layout of the RF test cleaning chamber (GV10x P3: ICP plasma; OSP: optical spectrometer; QCM: quartz crystal monitor; CDG: capacitive diaphragm gauge).

A comparative C1s XPS analysis of highly oriented pyrolytic graphite (HOPG - as a reference for pure $sp^2$ C), the amorphous carbon sample (*Amorphous C*), and the DLC sample (as a reference for $sp^3$ C) are shown in **Fig. 2**. The C1s XPS peak for *DLC* sample is up-shifted by 1.1 eV higher binding energies (BEs) with respect to the C1s line for the HOPG reference sample (284.6 eV), and is thus observed at 285.7 eV BE. This chemical shift gives clear evidence for the $sp^3$ configuration of the *DLC* sample [8]. On the other hand, the C1s line from amorphous C thin films exhibits a broad line shape that is composed of both the C1s lines for $sp^2$ as well as $sp^3$ C, thus giving additional spectroscopic evidence (combined with the Raman results given below) for the amorphous C character of that custom-made carbon thin film sample.

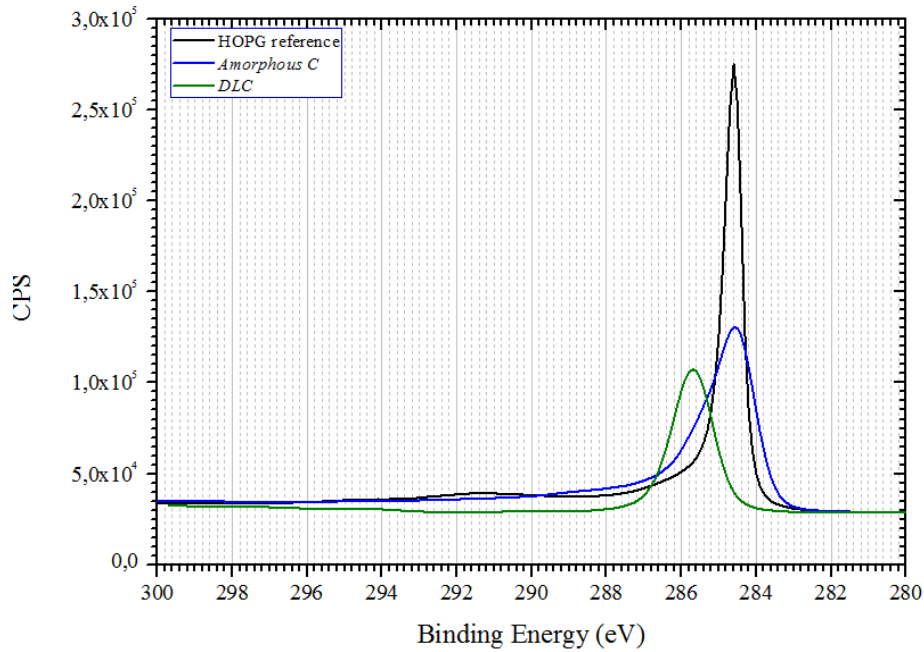

**Fig. 2** C1s XPS spectra of the HOPG reference sample and the diamond-like sample (*DLC*).

All samples were also characterized by Raman spectroscopy (see **Fig. 3(b)**). The prominent G line (1583 cm$^{-1}$) of the HOPG reference sample corresponds to a high-frequency LO phonon within the graphitic planes ($E_{2g}$ phonon in **Fig. 3(a)**), and it is present in various graphitic structures [9]. Another common peak in graphitic structures, the D peak or defect peak (1345 cm$^{-1}$; neither shown nor present in HOPG), corresponds to a TO phonon around the K point and arises from the breathing modes of six-atom carbon rings (see **Fig. 3(a)**). The reason why we do not observe the latter in the HOPG reference Raman spectrum is due to the fact that it must be activated by boundary phonons which are not present in a defect-less graphitic lattice (i.e., in HOPG) as they do not satisfy the Raman fundamental selection rule [10]. Nevertheless, we do observe a very wide D peak for the *Amorphous C* sample (blue solid line), centred just below 1500 cm$^{-1}$ and convoluted with a broad peak width. The broadening effect that the D and G peaks undergo is due to the small cluster sizes and a further broadening due to inequivalent chemical bonding. The 2D peak (2690 cm$^{-1}$) corresponds to the D peak overtone (i.e., second order of boundary phonons) and, as it originates from a process where momentum conservation is satisfied by two phonons with opposite wave vectors (double-resonance), no defects are required for its activation, and it is thus always present in sp$^2$ carbon [11, 12].

The "Diamond" peak (at 1332 cm$^{-1}$) corresponding to the first-order Raman scattering mode, being characteristic of diamond [13] and assigned to the $T_{2g}$ zone centre mode of the cubic diamond phase, at first glance does not show up in the *Amorphous C* case either. However, for the *Amorphous C* sample this peak is considerably broadened and thus also contributes to the asymmetric broad band observed between 1000 and 1700 cm$^{-1}$ (blue solid line in **Fig. 3(a)**).

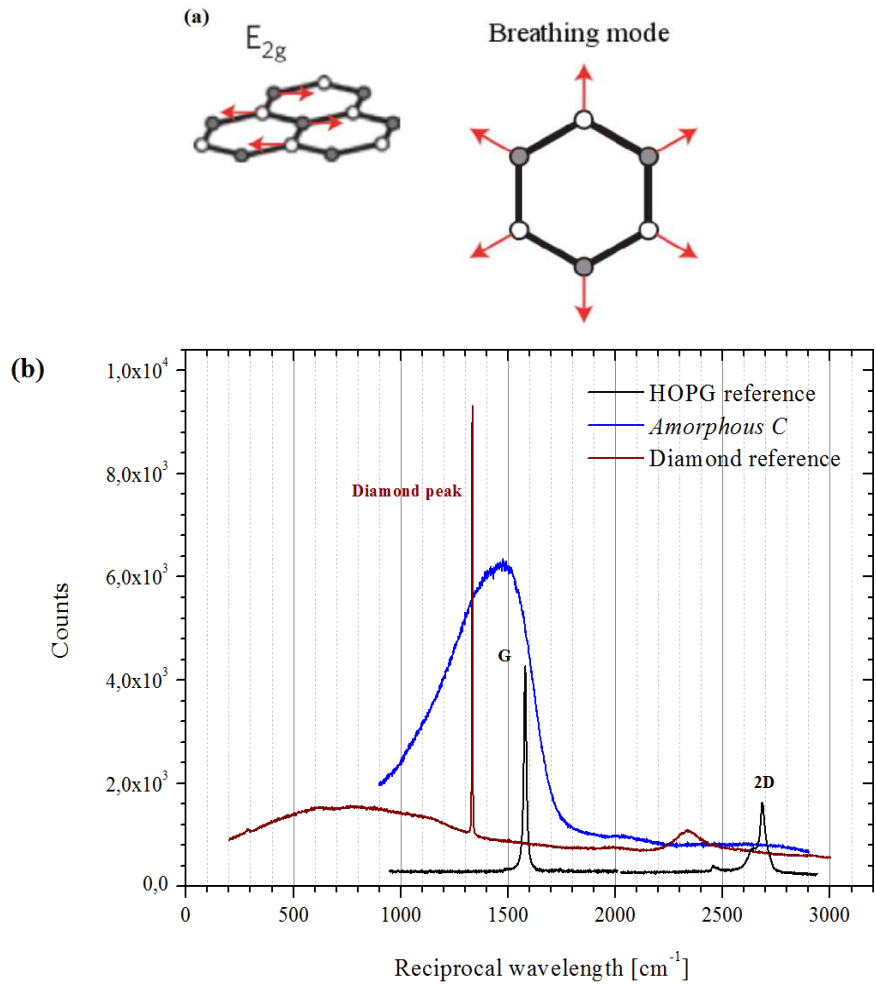

**Fig. 3(a)** Raman active phonon displacement pattern in graphitic planes. Empty and filled circles represent inequivalent carbon atoms. $E_{2g}$ give rise to the G peak whereas D peak partly arises from the "Breathing mode". **(b)** Raman spectra of the amorphous (*Amorphous C*) sample, DLC and HOPG reference samples.

Hence, the broad (blue) band for *Amorphous C* is composed of contributions from the widened D, G, and "Diamond" peaks, and thus corroborates a non-crystalline ordering among the $sp^2$ and $sp^3$ carbon atoms (i.e., a strong dominance of the amorphous phase).

## 3. RESULTS

Oxygen/argon and hydrogen/argon gas mixtures were used as feedstock gases for the plasma during the cleaning experiments, with a total pressure of $5.0 \times 10^{-3}$ mbar. These parameters were derived from the optimized concentrations and pressures (i.e., for highest cleaning rates) as found in our previous studies [1]: 95%/5% and 7%/93% for $O_2$/Ar and $H_2$/Ar, respectively. Relative concentrations were measured and systematically adjusted using the optical emission spectrum (OES) from the plasma as a reference. For the case of the $O_2$/Ar plasma, we focused onto the atomic lines for the neutral OI (or O•) radicals and the neutral ArI atoms, corresponding to the atomic transitions $3s^5S^0 - 3p^5P$ and $4s^2[1/2] - 4p^2[1/2]$, with a line strength of $3.69 \cdot 10^7$ s$^{-1}$ and $4.45 \cdot 10^7$ s$^{-1}$, respectively. In the case of $H_2$/Ar, we used the neutral HI (or H•) radicals ($4.41 \cdot 10^7$ s$^{-1}$) and the same line for ArI [14].

### 3.1 Cleaning experiments using $O_2$/Ar plasma

We first discuss the cleaning experiments performed on the *Amorphous C* samples using an $O_2$/Ar plasma. Here, we will focus onto the atomic lines for the neutral OI radicals and the neutral ArI atoms. The measured OES spectra from the $O_2$/Ar experiments for each RF power are shown in **Fig. 4(a)**. Please note the different vertical scales regarding the OES line intensities for each panel that show an increase of the OES lines concomitant with the increasing RF power. As can be seen from **Fig. 4(b)**, the intensity of the OI lines at 777.2 nm and 844.64 nm increase linearly as a function of RF power, indicating an increase of the density of OI species with RF power. The OES line related to the ArI emission at 750.39 nm remains at almost constant intensity for the whole range of RF power.

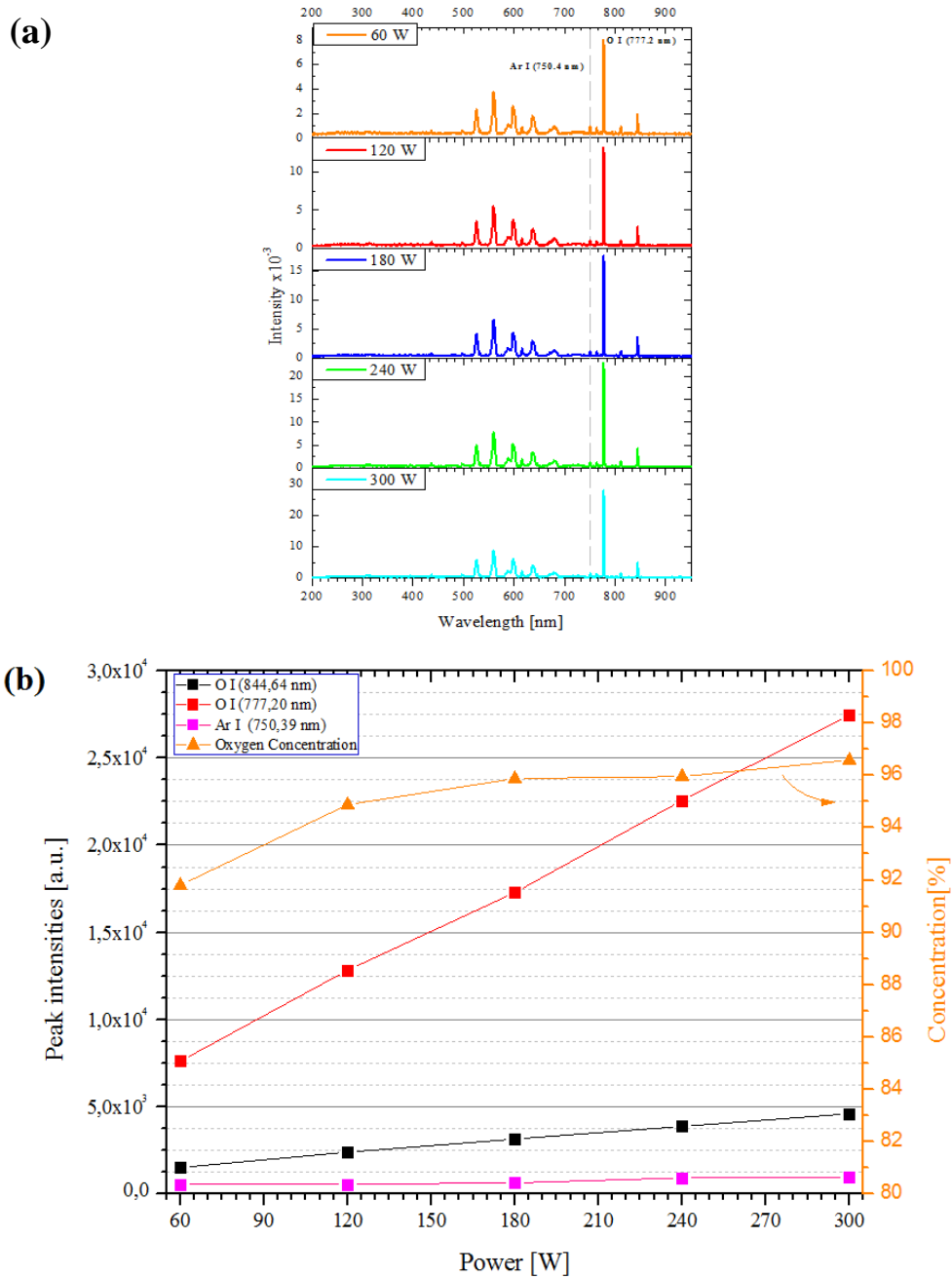

**Fig. 4(a)** Optical emission spectra from a typical $O_2$/Ar cleaning experiment. **(b)** Intensities of the principal emission lines from the atomic species within the plasma (left hand scale) as well as oxygen concentration (right hand scale) as a function of RF power. "O I" and "Ar I" refer to the left hand side axis (OES peak intensities) meanwhile "Oxygen Concentration" refers to the right hand side axis (Concentration).

**Fig. 5(a)** shows the decrease of the carbon layer thickness on both QCM sensors as a function of time and RF power as an additional parameter. As expected and as can be seen from the cleaning rates in **Fig. 5(b)** derived from **Fig. 5(a)**, we measured faster cleaning rates for the QCM sensor closer to the plasma source (QCM 1) as compared to the more distant sample on QCM2, so the quartz crystals on QCM1 became fully cleaned in a shorter amount of time. In the present case, the cleaning rates for both QCM sensors exhibit a linear dependence from the applied RF power which correlates with the linear increase of the OI OES lines in **Fig. 4(b)**. Data using both P3 and P4 plasma sources were acquired for our present case of $O_2$/Ar plasma cleaning of *Amorphous C* samples (as reported in *2.1 Experimental setup,* the P3 and P4 sources can reach 300W and 100W of RF power respectively, and P3 has a larger plasma cavity volume with respect to that of the P4 by a factor of 1.5).

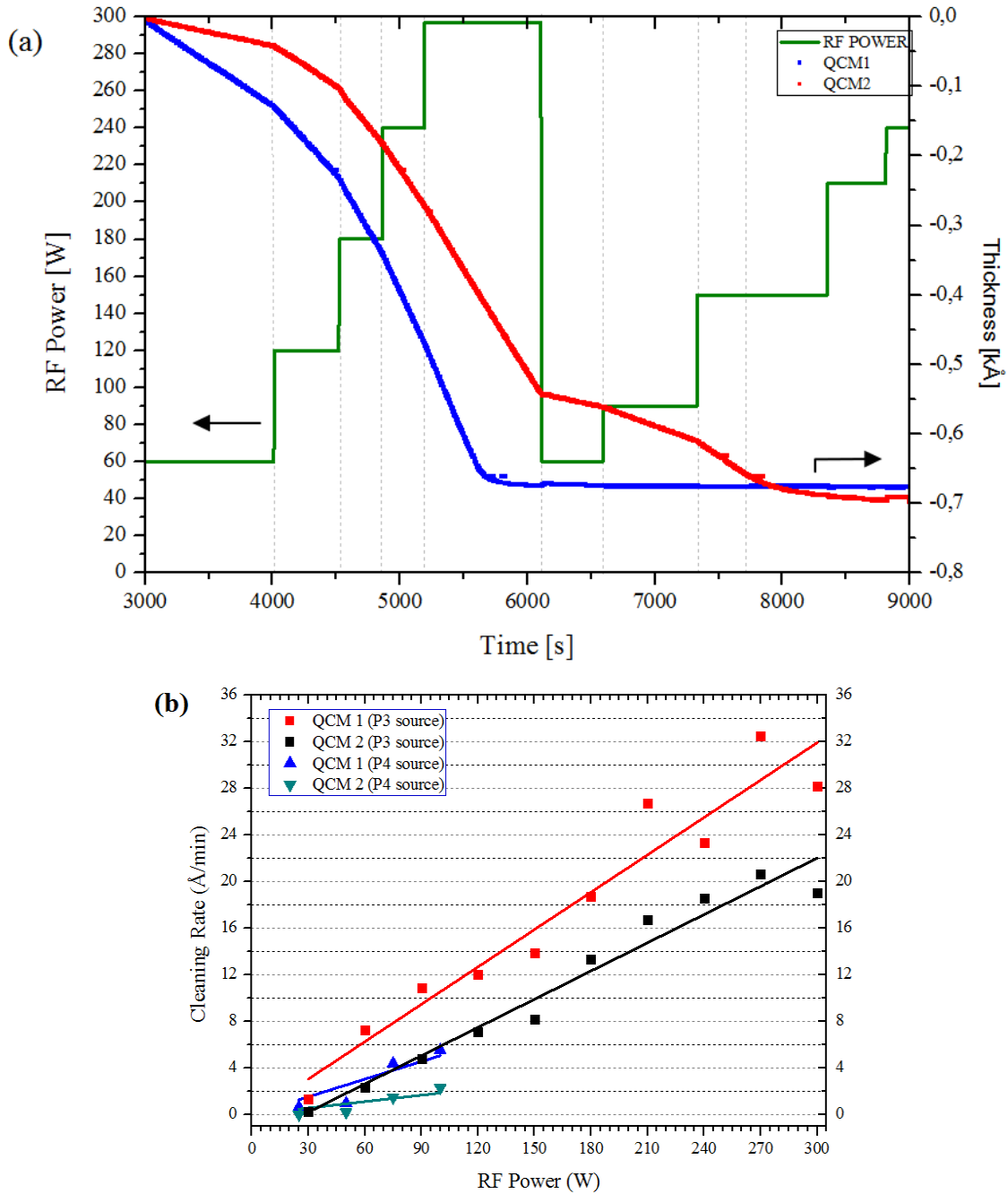

**Fig. 5(a)** Measured thickness decrease of the carbon layer on both QCM sensors (red and blue curves) and RF power (green curve) as a function of time. **(b)** Carbon cleaning rates as a function of RF Power, for the $O_2$/Ar cleaning of A*morphous C* samples. The lines represent linear fits to the data.

The results from the linear fits for the data from each individual QCM sensor in **Fig. 5(b),** for the case of the P3 source, are as follows:

$$C_R[\text{Å/min}] = 0{,}107 \pm 0.010 [\text{Å/min·W}]\, P_{RF}$$

for QCM1 (1.1)

$$C_R[\text{Å/min}] = 0{,}081 \pm 0.006 [\text{Å/min·W}]\, P_{RF}$$

for QCM2 (1.2)

where $C_R$ is the cleaning rate and $P_{RF}$ is the RF power. According to this, the cleaning rates decrease by a factor of roughly 1.3 with an increase in distance by roughly a factor of 2 (i.e., when going from 415 mm for QCM1 to 840 mm for QCM2).

In order to compare the performance of the P4 and P3 plasma sources with different plasma chamber volumes, we included the cleaning rates from a P4 plasma source with a plasma tube being smaller by a factor of 1.5 into **Fig5.(b)**, while using otherwise identical experimental setups and parameters. The results from the linear fits for the data from each individual QCM sensor in the case of the P4 source in **Fig. 5(b)** are as follows:

$$C_R[\text{Å/min}] = 0{,}050 \pm 0.007 [\text{Å/min·W}]\, P_{RF}$$

for QCM1 (1.3)

$$C_R[\text{Å/min}] = 0{,}018 \pm 0.004 [\text{Å/min·W}]\, P_{RF}$$

for QCM2 (1.4)

Thus, by increasing the upstream plasma chamber volume by a factor of 1.5, we observe an increase in cleaning rate by a factor of 2.1 and 4.5 for QCM1 (closer to the plasma source) and QCM2 (more distant from the plasma source), respectively. We attribute the increase of cleaning rate to the higher density of chemically active OI (or O$^\bullet$) species due to the larger plasma tube volume of the P3 plasma source.

### 3.2 Cleaning experiments using $H_2$/Ar plasma

We proceed as in the same manner as in the previous case for the cleaning experiments under $H_2$/Ar plasma. The measured OES spectra from the $H_2$/Ar experiments for each RF power are shown in **Fig. 6(a)**. Please note the different vertical scales regarding the OES line intensities for the different panels in **Fig. 6(b)**. We note that there is an intensity increase for the ArI as well as the HI emission lines with increasing RF power, with a decline in both intensities starting at RF powers of 250 W and beyond. This parallel behaviour of the Ar and H OES lines is in contrast to the corresponding findings from section 3.1 regarding the $O_2$/Ar feedstock gas mixture, where the OI line showed a strong increase with RF power whereas the ArI line remained constant. This hints towards different interactions between the various feedstock gases with a corresponding impact onto the cleaning mechanism.

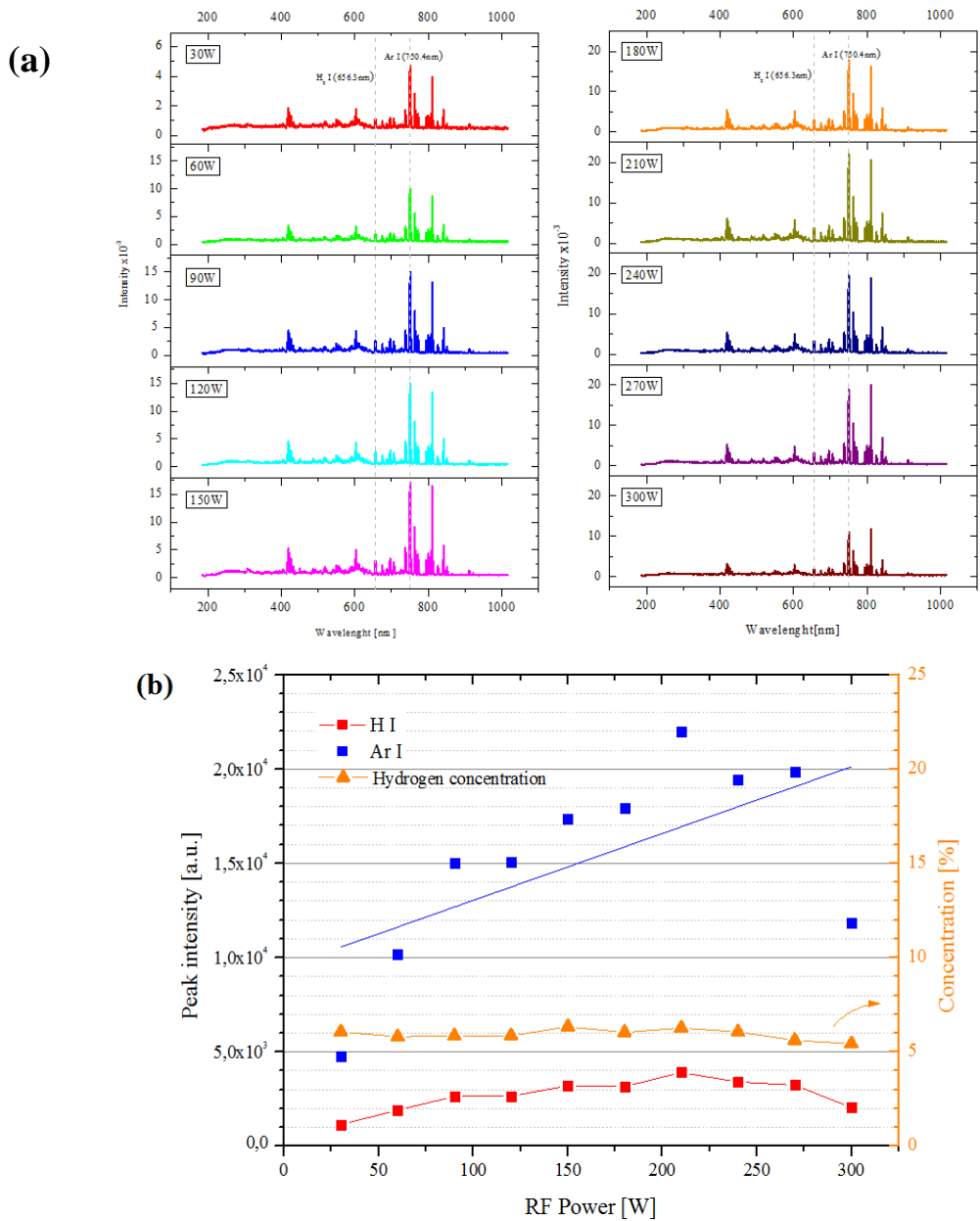

**Fig. 6(a)** Optical emission spectra from the H$_2$/Ar cleaning experiment. **(b)** Intensities of the principal emission lines from the atomic species within the plasma as a function of RF power. "H I" and "Ar I" refer to the left hand side axis (OES peak intensities) meanwhile "Hydrogen concentration" refers to the right hand side axis (Concentration).

**Fig. 7(a)** shows the decrease of the carbon layer thickness on both QCM sensors as a function of time and RF power as an additional parameter. As a first significant difference, we note the strongly non-linear behaviour of the cleaning rates shown in **Fig. 7(b)** with respect to the RF power, which exhibit a saturation starting at about 250 W. Regarding the absolute cleaning values, it is obvious that the cleaning obtained from the H$_2$/Ar plasma are about a factor of about 6 to 7 lower as compared to the O$_2$/Ar cleaning rates as has been observed previously [1]. In the present case of the H$_2$/Ar plasma, we thus also do observe a correlation between the applied RF power and the OES line intensities on one hand as well as the cleaning rates on the other hand, albeit the saturation in both is in sharp contrast to the corresponding linear behaviour in the case of the O$_2$/Ar plasma.

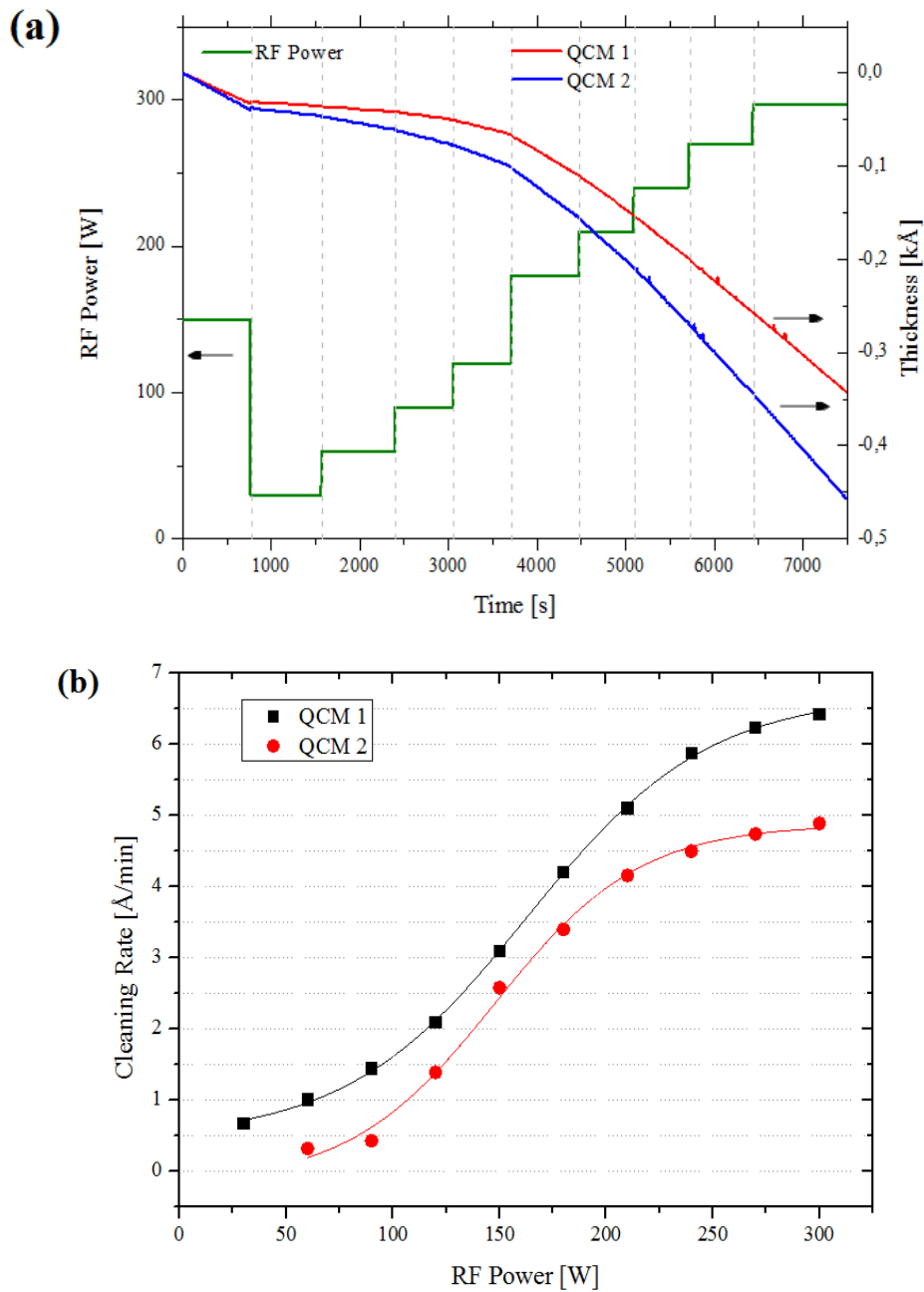

**Fig. 7(a)** Measured thickness decrease of the carbon layer on both QCM sensors (red and blue curves) and RF power (green curve) as a function of time. **(b)** Carbon cleaning rates as a function of RF Power for the $H_2/Ar$ cleaning of the A*morphous C* samples. The lines serve as guides to the eye.

3.3 Comparative cleaning experiments on a diamond-like carbon and amorphous carbon

In this study, the *Amorphous C* samples have been replaced with *DLC* samples by installing them into the *QCM 1* position (i.e., close to the plasma source). The procedure is the same as in the prior cases, so that the cleaning rates are obtained by determining the temporal derivative of the removed carbon thickness as measured by the QCM quartz crystal balances. **Fig. 8** shows the cleaning rate dependence from the RF power (limited to 100 W) for experiments using $O_2/Ar$ or $H_2/Ar$ plasma, respectively.

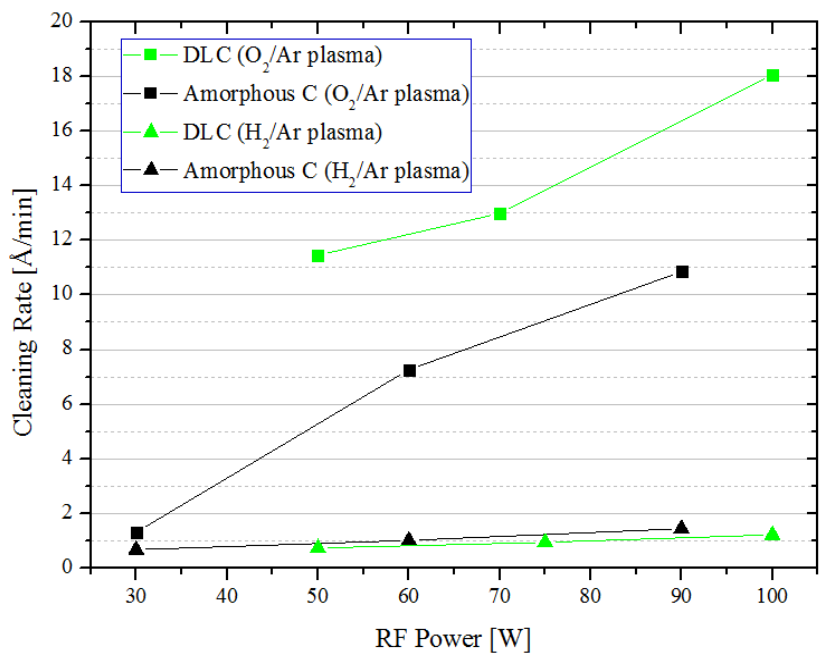

**Fig. 8** Comparison between the cleaning rates of *Amorphous C* and DLC samples, under $O_2$/Ar plasma (green and black lines above), and under $H_2$/Ar plasma (black and green lines below). All of them measured in QCM 1.

The cleaning rates for the *DLC* samples are higher for the $O_2$/Ar case with respect to the values found in the *Amorphous C* samples, meanwhile values for the *DLC* and *Amorphous C* turn out to be comparable when cleaning with $H_2$/Ar plasma. Hence, we can already anticipate a conclusion: $sp^3$ - hybridized carbon atoms are more reactive than $sp^2$ - hybridized carbon atoms with respect to the oxygen radicals within the $O_2$/Ar plasma.

## 4. DISCUSSION

After analysing the complete datasets, the results turn out to be consistent with respect to each other. The carbon removal rates increase with RF power and they decrease when moving away from the plasma source for both $O_2$/Ar and $H_2$/Ar plasma. Table I gives an overview on the increase in carbon cleaning rates as well as in the corresponding OES line intensities for the RF power range between 60 and 210 W in which all experiments show a linear behaviour as a function of RF power.

As can be seen from **Fig. 4(b)** and **5(b)**, for the case of the $O_2$/Ar plasma cleaning both the OI OES lines intensities as well as the carbon cleaning rate show a linear behaviour for the full RF power range up to 300 W. This result can be expected since the increase of RF power promotes the creation of chemically active species such as, e.g., oxygen radicals in the plasma that are responsible for the removal of the deposited carbon layer. Interestingly, the QCM1 cleaning rate for the $O_2$/Ar cleaning shown in **Table I** scales with almost the same factor as the OI OES line intensity (i.e., 3.5 versus 3.7), indicating a one-to-one relationship between the OI radical generation and the carbon removal process.

| | RF Power | QCM1 cleaning rate (Å/min) | QCM2 cleaning rate (Å/min) | Decrease with QCM distance | OI OES line increase at 777.2 nm | ArI OES line increase at 750.4 nm | HI OES line increase at 656.3 nm |
|---|---|---|---|---|---|---|---|
| **$O_2$/Ar cleaning** | 120 W | 22 | 16.5 | x $(1.3)^{-1}$ | | | |
| | 60 W | 6 | 3 | x $(2.0)^{-1}$ | | | |
| | *Increase with RF Power* | x 3.5 | x 5.5 | | x 3.7 | x 1.7 | -- |
| **$H_2$/Ar cleaning** | 120 W | 5.7 | 4.2 | x $(1.3)^{-1}$ | | | |
| | 60 W | 1.1 | 0.3 | x $(3.7)^{-1}$ | | | |
| | *Increase with RF Power* | x 5.2 | x 14 | | -- | x 2 | x 1.5 |

**Table I:** Increase/decrease factors for the carbon cleaning rates and the corresponding OES line intensities, as measured at 60 and at 210 W RF power (i.e., within the range of linear increase).

We differentiate between two different attenuation mechanisms for neutral radicals (i.e., cleaning agents) while traveling within the plasma along the longitudinal axis of the cylindrical chamber (i.e., emerging from the point-like plasma source and reaching sensor QCM1 first, and then sensor QCM2): Diffusion/recombination and geometrical dilution. Non-relativistic particles (such as oxygen and hydrogen radicals in our case) have a finite probability to collide within the plasma and hence to recombine, thus losing their chemical reactivity for cleaning purposes. This probability depends on the radical's energy and reactivity, where the former depends on the applied RF power. For a given energy (let us assume radicals with the same thermal energy at room temperature) and density, oxygen radicals will have a smaller reactivity than hydrogen radicals, thus leading to an enhanced probability for the latter to recombine (i.e., a smaller inelastic mean free path). Diffusion coefficients will therefore be larger for oxygen than for hydrogen radicals (we note that the similarity parameter $D \cdot p$ for $O_2$ and Ar as carrier gases – where $D$ is the diffusion coefficient for ions and electrons and $p$ is the pressure for a specific carrier gas - are very close to each other [15]). On the other hand, the (purely geometrical) dilution mechanism of radicals – which scales with respect to the distance r from the source proportional to $r^{-3}$ - is also responsible for the attenuation of the radical density with distance, as the latter spread over a 2π solid angle once the radicals have left the point-like plasma source and propagate along the chamber towards the detectors.

We first focus on the RF power ranges between 60 and 210 W given in **Table I**, where no saturation phenomena do occur in the cleaning rates as a function of RF power.

For the case of the $O_2$/Ar plasma, cleaning rates decrease with the QCM1 to QCM2 distance by a factor of 1.3 and 2.0 at 120 W and 60 W, respectively, whereas in the case of $H_2$/Ar we find a reduction factor of 1.3 at 120 W and a surprisingly large factor of 3.7 at 60 W (as shown in **Table I**). Assuming that the geometrical dilution mechanism is independent from RF power, this large reduction by a factor of 3.7 speaks in favour of a strong contribution by the diffusion/recombination mechanism. We hypothesize that the low 60W RF power leads to a low density of radicals, thus allowing for a larger mean free path length and hence making the diffusion mechanism more sensitive to distances within the range of the QCM1 and QCM2 sensors, as the large reduction of factor of 3.7 points out. However, at the conditions of a higher density of radicals right at the source exit (i.e., at 120 W), most of the recombination processes of the hydrogen radicals already take place well before reaching sensor QCM1, thus minimizing the differential effect between sensors QCM1 and QCM2 (decrease factor of 1.3). See **Fig. 9** for an illustration of these two different scenarios.

This effect is better observed for the H₂/Ar case than in the O₂/Ar case, because of the lower diffusivity for hydrogen than for oxygen radicals. Thus, for both the 120W and 60W cases using an O₂/Ar plasma, there is not much appreciable difference with the RF power variation between QCM1 and QCM2, with similar reduction factors of 1.3 (at 120 W) and 2.0 (at 60 W), respectively. Further measurements of the ions/electrons energy distribution with, e.g., a Faraday cup at different distances and powers, would help to corroborate this hypothesis and get a more quantitative picture for **Fig. 9**.

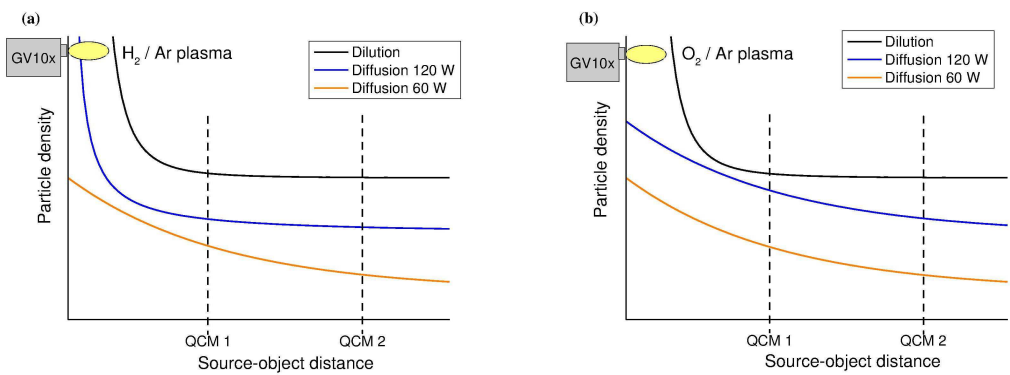

**Fig. 9:** Qualitative comparison between the dilution and diffusion mechanisms at 60 W and 120 W, for the case of (a) H₂/Ar and (b) O₂/Ar plasma. For the H₂/Ar plasma, even though the absolute number of charged ions is larger at 120 W, the diffusion rate has already reached a "flat" region when passing by the QCM sensors, so that the cleaning rates are not as power-dependent as for the 60W low RF power case. This does not apply to the O₂/Ar case.

Focusing now on the increase of cleaning rates with RF power, one can see in **Table I** that for the case of H₂/Ar plasma, the pertinent factors are 5.2 and 14 for the QCM1 and QCM2 sensors, respectively. Besides an increase of HI radical density due to the larger RF power, the larger increase in cleaning rate for the case of the more distant QCM2 sensor together with the much smaller increase for the HI OES line (factor of 1.5) indicates a combined/cooperative effect of the hydrogen radicals and the increase of UV light generated by the Ar species regarding the carbon removal via a photon-assisted breaking of carbon bonds and the subsequent reduction of carbon to hydrocarbon gas [16].

On the other hand and as already mentioned above, the increase by a factor of 3.5 and 5.5 for the QCM1 and QCM2 sensors for the O₂/Ar plasma as a function of RF power together with a similarly strong increase of the OI OES line by a factor of 3.7 yields a one-to-one relationship between the density of OI radicals and cleaning rates. The larger increase for the more distant QCM2 indicates an additional contribution by the inelastic mean free path length of the OI radicals at source-QCM2 distances. All this demonstrates that in the specific case of O₂/Ar feedstock gases the application of larger RF powers are beneficiary for larger optical components to be cleaned due to (i) a higher density of reactive OI radicals and (ii) a more effective inelastic mean free path length at larger source-object distances by dilution.

We now discuss the power dependence of the cleaning rates in **Fig. 5(b)** and **7(b)** up to the maximum RF power of 300 W. In contrast to the linear increase of the O₂/Ar cleaning rates with power as discussed above, the behaviour of the cleaning rates in the case of the H₂/Ar plasma is completely different: At lower RF power the cleaning rates grow exponentially, but at higher RF power (i.e., at 250 W and beyond), there is a distinct saturation of the carbon removal rate. Also, as can be seen from **Fig. 6(b)** and **Table I**, the number of HI radicals (as given by the intensity of the HI OES line) as well as the Ar I OES line show a rather weak linear increase by a factor of about 1.5 and 2, respectively, up to about 210 W RF power. This is in contrast to the large increase in QCM1 and QCM2 cleaning rates by a factor of 5.2 and 14, respectively. Again, the larger increase in cleaning rate - as compared to the smaller increase for the OES lines – could indicate a combined/cooperative effect of the hydrogen radicals and the increase of UV light generated by the Ar for the carbon removal.

Regarding the saturation in terms of cleaning rate for the specific case of the $H_2$/Ar plasma at high RF power (see **Fig. 7(b)**), we assume that this is not caused by a passivation of the sample surface by the plasma, as we do not observe this type of saturation effect for the $O_2$/Ar case. As discussed in ref. [1], we rather attribute this saturation to the recombination of hydrogen HI radicals into $H_2$ molecules by their mutual interaction and/or via the interaction with vacuum recipient walls as has already been discussed by several authors [17, 18]. This effect appears to be much more intense in the case of the hydrogen radicals as they have a higher reactivity than the oxygen OI radicals. Hence, the cleaning experiments using $H_2$/Ar suffer from saturation in the density of HI radicals available for the cleaning process, which explains the saturation at high RF powers in **Fig. 7(b)**. Indeed, we measured unusually elevated temperatures around $70^0$C downstream on the metal vacuum tubes of the plasma source, when operating it with $H_2$/Ar at high RF powers for a longer period of time. This effect occurs presumably due to the exothermic recombination of hydrogen radicals into molecules on the in-vacuum metal surfaces. All this fits well into the overall conceptual approach that hydrogen plasma need either to be embedded into an inert carrier gas matrix (such as, e.g., Ar) or to be provided with lower HI radical densities in order to allow for the mean free path lengths required for any large scale cleaning application. A direct measurement of the HI radical density at the QCM sensor is needed in order to corroborate the hypothesis of HI radical loss due to recombination.

We now discuss the carbon cleaning efficiency for the two different carbon allotropes under investigation (see **Fig. 8**). The differences between the cleaning rates for *Amorphous C* and *DLC* samples were studied separately for the case of $O_2$/Ar and $H_2$/Ar plasma. The $sp^3$ carbon (*DLC* samples) turned out to exhibit cleaning rates about 1.4 times higher than the amorphous samples (*Amorphous C*) under $O_2$/Ar plasma, which speaks in favour of an enhanced reactivity of the $sp^3$-bonded carbon atoms as compared to $sp^2$ carbon. On the other hand, there is no evidence for a significant change in cleaning speed for the two different allotropes in the case of a $H_2$/Ar plasma. Although the difference in cleaning speeds between *Amorphous C* and *DLC* for an $O_2$/Ar plasma is significant, we conclude that this latter feedstock gas combination is an appropriate choice for an efficient cleaning of both carbon allotropes, thus covering most optics cleaning applications. As mentioned in ref. [1], albeit the strongly reduced cleaning rates of $H_2$/Ar plasma, it is the method of choice for non-noble metal optical coatings that would be subject to oxidation by an $O_2$/Ar cleaning process.

Several early as well as more recent studies [19, 20, 21] aimed at the deposition of graphene nano-sheets, single-crystalline graphene, diamond or their nano-crystalline counterparts have emphasized the importance of adding low quantities of either oxygen or hydrogen to the main feedstock gases such as, e.g., methane or diethylene where the latter are used as carbon sources for the thin film deposition process. In this context, the mentioned hydrogen or oxygen add-on gases are then used as cleaning agents for avoiding the nucleation of undesirable $sp^2$ or $sp^3$ carbon intergrowths, depending on the specific type of carbon allotrope to be grown.

5. **CONCLUSIONS**

We have performed a comparative study regarding the low-pressure (i.e., $5x10^{-3}$ mbar) RF plasma cleaning rates for $O_2$/Ar and $H_2$/Ar feedstock gases throughout a wide range of RF powers and source-sample distances which are parameters being relevant for and compatible with applications in the field of cleaning of optical precision surfaces of real-size optics from carbon contaminations made of virtually different carbon allotropes. From this study, we derive the following conclusions:

- Satisfactory cleaning rates well beyond 10 Å per minute can be obtained in the case of reasonably large optics (i.e., of 1 m length or more) using $O_2$/Ar feedstock gases with RF powers just beyond 150 W.

- The above holds for both amorphous carbon (i.e., $sp^2$ and $sp^3$ mixtures) and especially diamond-like carbon $sp^3$ allotropes.

- The carbon cleaning rates for $O_2$/Ar feedstock gas as well as the density of OI (or O$^\cdot$) neutral oxygen radicals scale linearly with increasing RF power. The increase in cleaning rate as a function of RF power increases with increasing source-object distance, thus indicating the influence of a less effective OI inelastic mean free path length for larger source-object distances.

- The carbon cleaning rates for $H_2$/Ar feedstock gas exhibits a strongly "S-shaped" sigmoidal behaviour with increasing RF power saturating at higher powers, whereas the density of HI neutral radicals shows a (weak) linear increase with RF power. From this, we conclude a combined/cooperative effect between the HI (or H$^\bullet$) radicals and the UV light generated by the Ar atomic species regarding the carbon cleaning rates.

- For $H_2$/Ar plasma, the increase in cleaning rate as a function of RF power increases with increasing source-object distance (up to intermediate RF power), again indicating the influence of a less effective inelastic mean free path length for increasing source-object distances also for HI (or H$^\bullet$) radicals.

- At elevated RF powers and thus higher HI (or H$^\bullet$) radical densities, the high reactivity of the hydrogen radicals leads to an increase in recombination to chemically inactive $H_2$ molecules and thus to a saturation in cleaning rate.

**AKNOWLEDGEMENTS**

The authors would like to acknowledge the help by Guillaume Sauthier (ICN2) for his collaboration during the XPS characterization. We also thank ibss Group Inc. for kindly supplying the GV10x plasma sources as well as the DLC samples.